\documentclass[aps,pre,reprint,unsortedaddress,showpacs,amsmath,amssymb,twocolumn,preprintnumbers]{revtex4-1}
\usepackage{graphics,graphicx}

\begin{document}
\preprint{Physical Review E}

\title{Log-normal distribution from a process that is not multiplicative but is additive}

\author{Hideaki Mouri}
\affiliation{Meteorological Research Institute, Nagamine, Tsukuba 305-0052, Japan}

\date{October 8, 2013}

%__________________________________________________________________________________

\begin{abstract}
The central limit theorem ensures that a sum of random variables tends to a Gaussian distribution as their total number tends to infinity. However, for a class of positive random variables, we find that the sum tends faster to a log-normal distribution. Although the sum tends eventually to a Gaussian distribution, the distribution of the sum is always close to a log-normal distribution rather than to any Gaussian distribution if the summands are numerous enough. This is in contrast to the current consensus that any log-normal distribution is due to a product of random variables, i.e., a multiplicative process, or equivalently to nonlinearity of the system. In fact, the log-normal distribution is also observable for a sum, i.e., an additive process that is typical of linear systems. We show conditions for such a sum, an analytical example, and an application to random scalar fields such as of turbulence.
\end{abstract}

\pacs{02.50.Cw, 05.40.-a}

\maketitle

%__________________________________________________________________________________
%                                                                                 %
%  To the publisher:                                                              %
%                                                                                 %
%  (1)                                                                            %
%  You may find the expression ``$10^0$'' etc.                                    %
%  Please do not replace it with, e.g., ``1''.                                    %
%                                                                                 %
%  (2)                                                                            %
%  You may also find plus signs, e.g., ``+0.00'' in the text.                     %
%  Please do not remove such plus signs.                                          %
%                                                                                 %
%  (3)                                                                            %
%  If you want to modify any of our figures (letter size, letter position, ...),  %
%  please let us know.                                                            %
%  We would modify the figure by ourselves,                                       %
%  in order to avoid possible problems occuring during the modification.          %
%                                                                                 %
%__________________________________________________________________________________

\section{Introduction} \label{S1}

Consider a random variable that takes positive values, $z > 0$. If its logarithm $\ln z$ obeys a Gaussian (normal) distribution, $z$ itself is said to obey a log-normal distribution \cite{ks77},
%___________________________________________________________
\begin{equation}
\label{eq1}
f(z) = \frac{1}{z \sqrt{2 \pi \kappa_2}} \exp \left[ - \frac{(\ln z - \kappa_1)^2}{2 \kappa_2} \right].
\end{equation}
%___________________________________________________________
Here, $\kappa_1 = \langle \ln z \rangle$ and $\kappa_2 = \langle (\ln z - \langle \ln z \rangle)^2 \rangle$ are the average and variance of $\ln z$. This distribution is positively skewed with a long tail on the side of $z > \langle z \rangle$, except for the case of $\kappa_2 \ll 1$ where the distribution is rather close to a non-skewed Gaussian distribution.

The log-normal distribution is important because it has been observed very often at least as a good approximation in the natural and the social sciences \cite{lwsa01}. In physics, examples include fragment size \cite{k41a}, crystal size \cite{f57}, wave transmittance in random media \cite{t61}, turbulence energy \cite{m09}, dissipation rate of turbulence energy \cite{k62}, durations of transient events \cite{cp72,mhlm94}, stellar mass \cite{ms79}, and cosmological density fluctuations \cite{h34}. Also for temporal fluctuations of some nonlinear systems \cite{o62,bc03}, a fundamental parameter is considered to vary slowly in a log-normal distribution.

The current consensus is that any log-normal distribution is due to a multiplicative process, i.e., a product of random variables \cite{g30}, $\prod_{n=1}^N z_n$. Its logarithm is a sum of random variables, $\sum_{n=1}^N \ln z_n$. This tends to a Gaussian distribution as $N \rightarrow \infty$, according to the central limit theorem \cite{ks77}, if those summands have finite variances, if none of the summands dominates the others, and also if the summands are not highly dependent on one another. It follows that $\prod_{n=1}^N z_n$ tends to a log-normal distribution. The multiplicative process is in turn attributed to nonlinearity of the system.

However, for a class of positive random variables, we observe that their sum becomes log-normal before it becomes Gaussian. That is, so far as the total number $N$ of the summands is finite and is large enough, the distribution of the sum is always close to a log-normal distribution rather than to any Gaussian distribution. Since this log-normal distribution tends to a Gaussian distribution as $N \rightarrow \infty$, our observation is yet in accordance with the central limit theorem.

Thus, log-normality could be also observed for a sum of random variables, i.e., an additive process being typical of linear systems that are divisible into independent subsystems. Such a possibility has not been studied, despite the existence of many studies on the log-normal distribution. We describe the general theory in Sec.~\ref{S2} and an analytical example in Sec.~\ref{S3}. They are applied to random scalar fields such as of turbulence in Sec.~\ref{S4}. We conclude with remarks in Sec.~\ref{S5}.

\section{General Theory} \label{S2}

The additive process is studied by using an average of positive random variables $z_n > 0$,
%___________________________________________________________
\begin{subequations}
\begin{equation}
\label{eq2a}
\bar{z}_N = \frac{1}{N} \sum_{n=1}^N z_n.
\end{equation}
%___________________________________________________________
Instead of the usual moments $\langle z^m \rangle$, we use the cumulants $\langle z^m \rangle_c = d^m \ln \langle \exp (i \xi z ) \rangle /d(i \xi)^m \vert_{\xi=0}$. They are $\langle z^2 \rangle_c = \langle (z-\langle z \rangle)^2 \rangle$, $\langle z^3 \rangle_c = \langle (z-\langle z \rangle)^3 \rangle$, $\langle z^4 \rangle_c = \langle (z-\langle z \rangle)^4 \rangle - 3 \langle (z-\langle z \rangle)^2 \rangle^2$, and so on \cite{ks77}. The cumulant for $m=2$ is just the variance.

For simplicity (see Sec.~\ref{S1}), we assume that the random variables $z_n$ are independent of one another and are identically distributed with the distribution of some random variable $z_{\ast}$. Their cumulants are all assumed to be finite. From these assumptions, since any cumulant of a sum of independent random variables is the sum of cumulants of the variables \cite{ks77}, we have cumulants of $\bar{z}_N$ as
%___________________________________________________________
\begin{equation}
\label{eq2b}
\langle \bar{z}_N^m \rangle_c = \frac{1}{N^m} \sum_{n=1}^N \langle z_n^m \rangle_c = \frac{\langle z_{\ast}^m \rangle_c}{N^{m-1}} .
\end{equation}
%___________________________________________________________
Here, $\langle z_{\ast}^m \rangle_c$ has been substituted into each of $\langle z_n^m \rangle_c$. The skewness of $\bar{z}_N$ is
%___________________________________________________________
\begin{equation}
\label{eq2c}
\frac{\langle \bar{z}_N^3 \rangle_c}{\langle \bar{z}_N^2 \rangle_c^{1.5}} 
= 
\frac{1}{N^{0.5}} \frac{\langle z_{\ast}^3 \rangle_c}{\langle z_{\ast}^2 \rangle_c^{1.5}} .
\end{equation}
%___________________________________________________________
The kurtosis of $\bar{z}_N$ is
%___________________________________________________________
\begin{equation}
\label{eq2d}
\frac{\langle \bar{z}_N^4 \rangle_c}{\langle \bar{z}_N^2 \rangle_c^2} 
= 
\frac{1}{N} \frac{\langle z_{\ast}^4 \rangle_c}{\langle z_{\ast}^2 \rangle_c^2} .
\end{equation}
\end{subequations}
%___________________________________________________________
With an increase in $N$, these skewness and kurtosis decay to the Gaussian value of $0$. A faster decay to the Gaussian value of $0$ is obtained for $\langle \bar{z}_N^m \rangle_c / \langle \bar{z}_N^2 \rangle_c^{m/2} \propto 1/N^{m/2-1}$ at $m \ge 5$. Thus, $\bar{z}_N$ tends to a Gaussian distribution. This is the central limit theorem \cite{ks77}.

However, the above discussion does not determine the shapes of the far tails of the distribution at deviations of $\bar{z}_N$ from $\langle \bar{z}_N \rangle$ that are larger than several of $\langle \bar{z}_N^2 \rangle_c^{0.5}$. They are ignored in the central limit theorem \cite{c38,t09}. In fact, $\bar{z}_N > 0$ is by definition not exactly Gaussian. The theorem is nevertheless of very practical use because the shapes of those tails are not reliably determined with any finite size of data obtained from any actual observation. Also in any actual system, $N$ is finite. We define that $\bar{z}_N$ is observed to be Gaussian, regardless of its exact distribution, if its skewness and kurtosis are close enough to $0$.

The additive process of Eq.~(\ref{eq2a}) is regarded as if it were multiplicative. For $z_n = \langle z_n \rangle (1 + \varepsilon_n ) > 0$, we consider $N$ such that $N \gg \langle \varepsilon_n^2 \rangle_c^{0.5} = \langle z_{\ast}^2 \rangle_c^{0.5} / \langle z_{\ast} \rangle$. Then, large deviations of $\varepsilon_n$ from $\langle \varepsilon_n \rangle = 0$ are ignored to focus on $\varepsilon_n$ within the range of $-N \ll \varepsilon_n \ll +N$,
%___________________________________________________________
\begin{subequations}
\label{eq3}
\begin{equation}
\label{eq3a}
\frac{1}{N} \sum_{n=1}^N (1+\varepsilon_n )
= 
1+\frac{\varepsilon_1}{N}+ ... +\frac{\varepsilon_N}{N}
\simeq 
\left[ \prod_{n=1}^N (1+\varepsilon_n) \right]^{1/N}.
\end{equation}
%___________________________________________________________
The arithmetic average of $z_n$ is thereby approximated as their geometric average,
%___________________________________________________________
\begin{equation}
\label{eq3b}
\frac{1}{N} \sum_{n=1}^N z_n \simeq \left( \prod_{n=1}^N z_n \right)^{1/N}.
\end{equation}
\end{subequations}
%___________________________________________________________
From Eqs.~(\ref{eq2a}) and (\ref{eq3b}), we obtain an approximation as a multiplicative process
%___________________________________________________________
\begin{subequations}
\label{eq4}
\begin{equation}
\label{eq4a}
\bar{z}_N \simeq \left( \prod_{n=1}^N z_n \right)^{1/N}
\ \
\mbox{or} 
\ \
\ln \bar{z}_N \simeq \frac{1}{N} \sum_{n=1}^N \ln z_n .
\end{equation}
%___________________________________________________________
This is to be used as an exact relation. Indeed, if $z_n$ were distributed only within a finite range, Eq.~(\ref{eq4a}) would be exact in the limit $N \rightarrow \infty$. The cumulants of $\ln \bar{z}_N$ are
%___________________________________________________________
\begin{equation}
\label{eq4b}
\langle (\ln \bar{z}_N)^m \rangle_c 
= 
\frac{1}{N^m} \sum_{n=1}^N \langle (\ln z_n)^m \rangle_c 
= 
\frac{\langle (\ln z_{\ast})^m \rangle_c}{N^{m-1}} .
\end{equation}
%___________________________________________________________
The skewness and kurtosis of $\ln \bar{z}_N$ are
%___________________________________________________________
\begin{equation}
\label{eq4c}
\frac{\langle (\ln \bar{z}_N)^3 \rangle_c}{\langle (\ln \bar{z}_N)^2 \rangle_c^{1.5}} 
= 
\frac{1}{N^{0.5}} \frac{\langle (\ln z_{\ast})^3 \rangle_c}{\langle (\ln z_{\ast})^2 \rangle_c^{1.5}} ,
\end{equation}
%___________________________________________________________
and
%___________________________________________________________
\begin{equation}
\label{eq4d}
\frac{\langle (\ln \bar{z}_N)^4 \rangle_c}{\langle (\ln \bar{z}_N)^2 \rangle_c^2} 
= 
\frac{1}{N} \frac{\langle (\ln z_{\ast})^4 \rangle_c}{\langle (\ln z_{\ast})^2 \rangle_c^2} .
\end{equation}
\end{subequations}
%___________________________________________________________
With an increase in $N$, they decay to the Gaussian value of $0$. Thus, although $\bar{z}_N$ is observed to become Gaussian, $\bar{z}_N$ is also observed to become log-normal.

The above observations are approximate in that they have ignored the large deviations of $\bar{z}_N$ from $\langle \bar{z}_N \rangle$. Those ignored to observe the log-normal distribution are not necessarily the same as those ignored to observe the Gaussian distribution. Since the log-normal distribution has to obey asymptotes such as $\langle z^3 \rangle_c / \langle z^2 \rangle_c^{1.5} = 3 \langle (\ln z)^2 \rangle_c^{0.5}$ and $\langle z^4 \rangle_c / \langle z^2 \rangle_c^2 = 16 \langle (\ln z)^2 \rangle_c$ at $\langle (\ln z)^2 \rangle_c \ll 1$ \cite{ks77}, it tends eventually to the Gaussian distribution as $N \rightarrow \infty$ and hence as $\langle (\ln \bar{z}_N)^2 \rangle_c \rightarrow 0$, in accordance with the central limit theorem.

The large deviations have been also ignored in Eqs.~(\ref{eq3}) and (\ref{eq4}), but the resultant log-normality is not necessarily less accurate than the Gaussianity, which results from a similar approximation in the central limit theorem. We also note that Eqs.~(\ref{eq3}) and (\ref{eq4}) could become much more accurate if the same constant were added to each of $z_n$ so that $\langle \varepsilon_n^2 \rangle_c^{0.5}$ were smaller (see Sec.~\ref{S4}).

For $\bar{z}_N$ to tend to a log-normal distribution faster than to any Gaussian distribution, necessary conditions are obtained by comparing Eqs.~(\ref{eq4c}) and (\ref{eq4d}) with Eqs.~(\ref{eq2c}) and (\ref{eq2d}), 
%___________________________________________________________
\begin{subequations}
\label{eq5}
\begin{equation}
\label{eq5a}
\left| \frac{\langle (\ln z_{\ast})^3 \rangle_c}{\langle (\ln z_{\ast})^2 \rangle_c^{1.5}} \right|
< 
\left| \frac{\langle z_{\ast}^3 \rangle_c}{\langle z_{\ast}^2 \rangle_c^{1.5}} \right| ,
\end{equation}
%___________________________________________________________
and
%___________________________________________________________
\begin{equation}
\label{eq5b}
\left| \frac{\langle (\ln z_{\ast})^4 \rangle_c}{\langle (\ln z_{\ast})^2 \rangle_c^2} \right|
< 
\left| \frac{\langle z_{\ast}^4 \rangle_c}{\langle z_{\ast}^2 \rangle_c^2} \right| .
\end{equation}
\end{subequations}
%___________________________________________________________
These are also practically sufficient conditions because it follows that $\bar{z}_N$ at large $N$ is always observed to be log-normal rather than to be Gaussian so far as the observation is based on the skewness and kurtosis as in our definition. If not identical were distributions of the individual summands $z_n$, their typical values could be used for Eq.~(\ref{eq5}).

The conditions of Eq.~(\ref{eq5}) hold for an extensive class of positively skewed distributions. It is actually known that the skewness of such a distribution, $\langle z^3 \rangle_c / \langle z^2 \rangle_c^{1.5} \gg 0$, is much reduced by transforming the variable $z$ into its logarithm $\ln z$ and thereby by reducing the positive tail \cite{ks77,j49}, i.e., $\langle (\ln z)^3 \rangle_c / \langle (\ln z)^2 \rangle_c^{1.5} \ll \langle z^3 \rangle_c / \langle z^2 \rangle_c^{1.5}$ but yet $\langle (\ln z)^3 \rangle_c / \langle (\ln z)^2 \rangle_c^{1.5} > - \langle z^3 \rangle_c / \langle z^2 \rangle_c^{1.5}$. The log-normal distribution is hence observable for some extensive class of additive processes. Examples are shown in the sections below.

\section{Analytical Example} \label{S3}
 
\begin{figure}[tbp]
\resizebox{8.8cm}{!}{\includegraphics*[4.9cm,7.8cm][17cm,26cm]{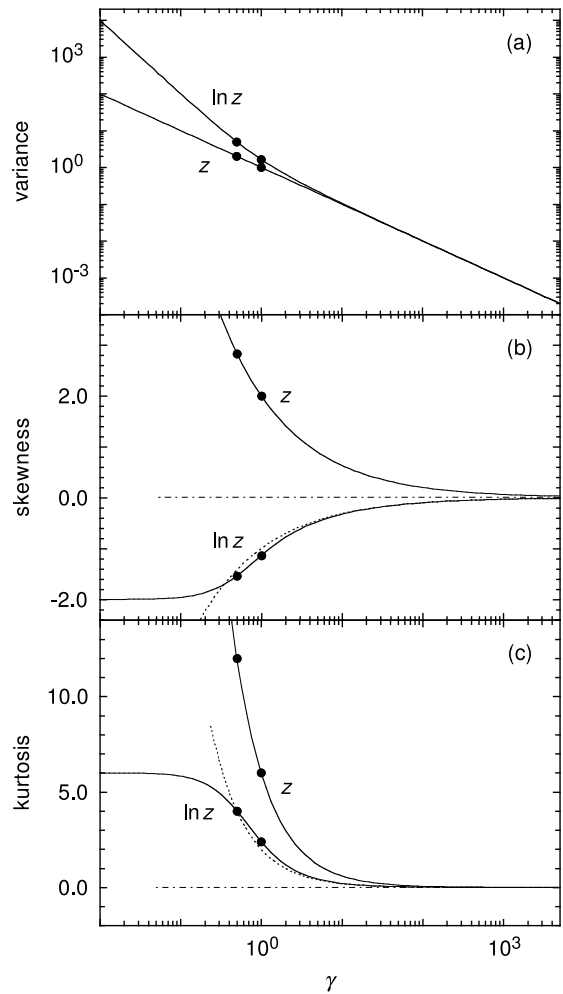}}
\caption{\label{f1} Variance (a), skewness (b), and kurtosis (c) of $z$ and of $\ln z$ as a function of $\gamma$ for the gamma distribution of Eq.~(\ref{eq6}). The variance of $z$ is normalized with the square of $\langle z \rangle$. We indicate the values at $\gamma = 1/2$ and $1$ with filled circles. The dotted curves are asymptotes of Eq.~(\ref{eq8c}). The dot-dashed lines denote the Gaussian value of $0$.}
\end{figure}

As an example to observe the log-normal distribution for an additive process, we study the gamma distribution of a positive random variable $z > 0$ \cite{ks77},
%___________________________________________________________
\begin{equation}
\label{eq6}
f(z) = \frac{1}{\Gamma (\gamma)} z^{\gamma-1} \exp(-z) 
\quad
\mbox{with}
\quad
\gamma > 0 .
\end{equation}
%___________________________________________________________
Here, $\Gamma$ is the gamma function. The parameter $\gamma$ defines the shape of the distribution, which is positively skewed especially when $\gamma$ is close to $0$. We have cumulants of $z$ as
%___________________________________________________________
\begin{subequations}
\label{eq7}
\begin{equation}
\label{eq7a}
\langle z^m \rangle_c = \gamma (m-1)! .
\end{equation}
%___________________________________________________________
Then, the skewness and kurtosis of $z$ are
%___________________________________________________________
\begin{equation}
\label{eq7b}
\frac{\langle z^3 \rangle_c}{\langle z^2 \rangle_c^{1.5}} = \frac{2}{\sqrt{\gamma}}
\quad
\mbox{and}
\quad
\frac{\langle z^4 \rangle_c}{\langle z^2 \rangle_c^2} = \frac{6}{\gamma} .
\end{equation}
\end{subequations}
%___________________________________________________________
For the cumulants of $\ln z$ of the gamma distribution, the definition $\langle (\ln z)^m \rangle_c = d^m \ln \langle \exp (i \xi \ln z ) \rangle /d(i \xi)^m \vert_{\xi=0}$ is known to yield an analytical formula \cite{ks77,j49},
%___________________________________________________________
\begin{subequations}
\label{eq8}
\begin{equation}
\label{eq8a}
\langle (\ln z)^m \rangle_c = \frac{d^m}{d \gamma^m} \ln \Gamma (\gamma) .
\end{equation}
%___________________________________________________________
The right-hand side is a poly-gamma function. By using its asymptote for $\gamma \rightarrow \infty$ \cite{as65}, we obtain
%___________________________________________________________
\begin{equation}
\label{eq8b}
\langle (\ln z)^m \rangle_c \rightarrow (-1)^m \frac{(m-2)!}{\gamma^{m-1}}  
\quad
\mbox{for}
\quad
m \ge 2.
\end{equation}
%___________________________________________________________
The skewness and kurtosis of $\ln z$ tend as
%___________________________________________________________
\begin{equation}
\label{eq8c}
\frac{\langle (\ln z)^3 \rangle_c}{\langle (\ln z)^2 \rangle_c^{1.5}} \rightarrow - \frac{1}{\sqrt{\gamma}} 
\quad
\mbox{and}
\quad
\frac{\langle (\ln z)^4 \rangle_c}{\langle (\ln z)^2 \rangle_c^2} \rightarrow \frac{2}{\gamma} .
\end{equation}
\end{subequations}
%___________________________________________________________
These asymptotes are good approximations of the skewness and kurtosis for $\gamma \gtrsim 10^0$, as shown in Fig.~\ref{f1} (dotted and solid curves).

%_________________________________________
%_________________________________________

The gamma distribution is reproductive \cite{ks77}. Recall the additive process of Eq.~(\ref{eq2a}). If its independent random variables $z_n$ are identically distributed with the gamma distribution for $\gamma = \gamma_{\ast}$, the functional form of Eq.~(\ref{eq7a}) ensures that $\sum_{n=1}^N z_n$ obeys the gamma distribution for $\gamma = N \gamma_{\ast}$. The shape is the same for the distribution of $\bar{z}_N = \sum_{n=1}^N z_n /N$.

With an increase in $N = \gamma / \gamma_{\ast}$, it follows from Eqs.~(\ref{eq7}) and (\ref{eq8}) for $m \ge 3$ that $\langle (\ln \bar{z}_N)^m \rangle_c / \langle (\ln \bar{z}_N)^2 \rangle_c^{m/2}$ decays to the Gaussian value of $0$ faster than $\langle \bar{z}_N^m \rangle_c / \langle \bar{z}_N^2 \rangle_c^{m/2}$ by a factor of $m-1$ in units of $N^{m/2-1}$. In particular, the skewness and kurtosis of $\ln \bar{z}_N$ decay faster than those of $\bar{z}_N$ by factors of $2$ and $3$ (see also Fig.~\ref{f1}). Thus, $\bar{z}_N$ tends to a log-normal distribution faster than to any Gaussian distribution. If $N$ is large enough, $\bar{z}_N$ is always observed to be log-normal, according to our definition of the observation in Sec.~\ref{S2}.

The gamma distribution arises from various processes in physics. Of importance are those for $\gamma_{\ast} = 1/2$ and $1$ (filled circles in Fig.~\ref{f1}). While $\gamma_{\ast} = 1/2$ corresponds to the distribution of the square of a zero-mean Gaussian random variable, $\gamma_{\ast} = 1$ corresponds to an exponential distribution. The gamma distribution is hence expected to explain some of observations of the log-normal distribution in terms of the additive process. For example, such an observation for event duration $\tau$ \cite{lwsa01,mhlm94,cp72} could be explained by a series of subevents, $\sum_{n=1}^N \tau_n$, if each of them is exponential, $f(\tau_n) \propto \exp(- \tau_n / \tau_{\ast} )$. Another example is the energy $E$ of a canonical ensemble in the statistical mechanics \cite{c85}, which is to be used in the Appendix.

The above features of the gamma distribution confirm our theory in Sec.~\ref{S2}. Being consistent with Eq.~(\ref{eq5}) that is satisfied for any of $\gamma_{\ast} > 0$, the distribution observed at any of $N = \gamma / \gamma_{\ast} \gg 1/ \gamma_{\ast}$ is log-normal rather than Gaussian (see Fig.~\ref{f1}). For $\gamma_{\ast} = 1/2$, we use Eq.~(\ref{eq8a}) to obtain the right-hand sides of Eqs.~(\ref{eq4c}) and (\ref{eq4d}) as $-1.54/N^{0.5}$ and $+4/N$. They are close to $-1.41/N^{0.5}$ and $+4/N$, the right-hand sides of Eq.~(\ref{eq8c}) for $\gamma = N \gamma_{\ast} = N/2$. A similar result is found for $\gamma_{\ast} =1$, where those of Eqs.~(\ref{eq4c}) and (\ref{eq4d}) are $-1.14/N^{0.5}$ and $+2.4/N$. Thus reliable are relations of Eq.~(\ref{eq4}), despite the fact that we have ignored the large deviations of $\bar{z}_N$ from $\langle \bar{z}_N \rangle$.

\section{Application to Random Field} \label{S4}

The log-normal distribution is to be observed for large-scale fluctuations of a random scalar field that is homogeneous along one-dimensional position $x$. Although we are interested in the energy $v^2(x)$ of a velocity component $v$ of turbulence as in our past works \cite{m09,m11}, the distribution also applies to, e.g., the cosmological density fluctuations \cite{Ba86}. If time $t$ is considered instead of the position $x$, the scalar is just the energy of some stochastic process \cite{r54}. The particular field studied here corresponds to the relaxed stationary state of the Ornstein-Uhlenbeck process that represents the Brownian motion and the Johnson noise \cite{g96a}.

To focus on the process for the log-normal distribution, we study an idealized Gaussian random field, where $v$ at each position $x$ obeys a zero-mean Gaussian distribution. Any statistical feature of such a field is determined by the two-point correlation of $v$ alone. Its functional form is set to be exponential,
%___________________________________________________________
\begin{subequations}
\label{eq9}
\begin{equation}
\label{eq9a}
\langle v(x+r) v(x) \rangle \propto \exp \left(- \frac{r}{L_v} \right) .
\end{equation}
%___________________________________________________________
Here, the average $\langle \cdot \rangle$ is taken over the position $x$. The two-point correlation of $v^2$ is
%___________________________________________________________
\begin{equation}
\label{eq9b}
\langle [v^2(x+r) - \langle v^2 \rangle ][ v^2(x) - \langle v^2 \rangle] \rangle 
\propto
\exp \left(- \frac{r}{L_{v^2}} \right) ,
\end{equation}
%___________________________________________________________
with the correlation length
%___________________________________________________________
\begin{equation}
\label{eq9c}
L_{v^2} = \frac{L_v}{2}.
\end{equation}
\end{subequations}
%___________________________________________________________
This field is numerically calculated by making use of a simple and exact algorithm for the Ornstein-Uhlenbeck process \cite{g96a,g96b}. Since its fluctuations exist at any small scale \cite{g96a}, the calculations are made with different resolutions if necessary.

The one-dimensional field is divided into segments with length $R$ \cite{m09,m11}. Over each of the segments, the center of which is tentatively defined as $x_{\ast}$, we average the energy $v^2$ as
%_________________________________________
\begin{equation}
\label{eq10}
\bar{v}_R^2(x_{\ast})=\frac{1}{R} \int^{+R/2}_{-R/2} v^2(x_{\ast}+x) \, dx .
\end{equation}
%_________________________________________
This corresponds to the additive process of Eq.~(\ref{eq2a}) in the case of $R \gg L_{v^2}$, where the correlation of $v^2$ is negligible. The segment serves as a linear system because it is divisible into independent subsegments \cite{m11}. Within the individual subsegments, we could regard $v^2$ as constants. Since the overall distribution of $v^2$ corresponds to Eq.~(\ref{eq6}) for $\gamma = 1/2$ and thereby satisfies Eq.~(\ref{eq5}) (see Sec.~\ref{S3}), a log-normal distribution could be observable for $\bar{v}_R^2$ among the segments with length $R \gg L_{v^2}$.

\begin{figure}[bp]
\resizebox{8.8cm}{!}{\includegraphics*[4.9cm,11.5cm][17cm,26cm]{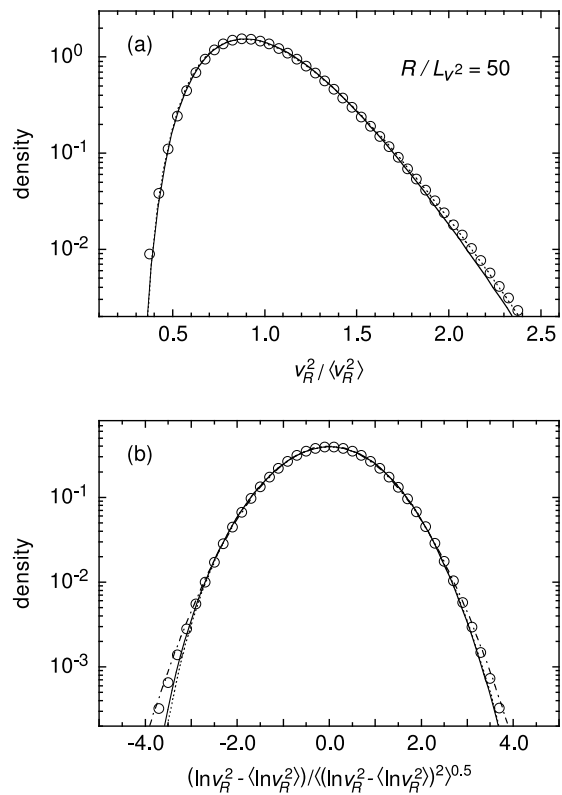}}
\caption{\label{f2} Probability density distributions of $\bar{v}^2_R / \langle \bar{v}^2_R \rangle $ (a) and of $( \ln \bar{v}^2_R - \langle \ln \bar{v}^2_R \rangle ) / \langle ( \ln \bar{v}^2_R - \langle \ln \bar{v}^2_R \rangle )^2 \rangle^{0.5}$ (b) at $R = 50 L_{v^2}$ for the random field of Eq.~(\ref{eq9}). The solid and the dotted curves are model predictions described in Sec.~\ref{S4} and in the Appendix. The dot-dashed curve denotes the Gaussian distribution.}
\end{figure}

Figure \ref{f2} compares the distribution of $\bar{v}_R^2$ with that of $\ln \bar{v}_R^2$ at $R = 50 L_{v^2}$ (open circles). While the distribution of $\bar{v}_R^2$ is positively skewed, that of $\ln \bar{v}_R^2$ is symmetric and is close to a Gaussian distribution (dot-dashed curve). Thus, $\bar{v}_R^2$ is log-normal at least as a good approximation, which is also known in terms of the Ornstein-Uhlenbeck process \cite{mt}.

%_________________________________________
%_________________________________________
\begin{figure}[tbp]
\resizebox{8.8cm}{!}{\includegraphics*[4.9cm,7.8cm][17cm,26cm]{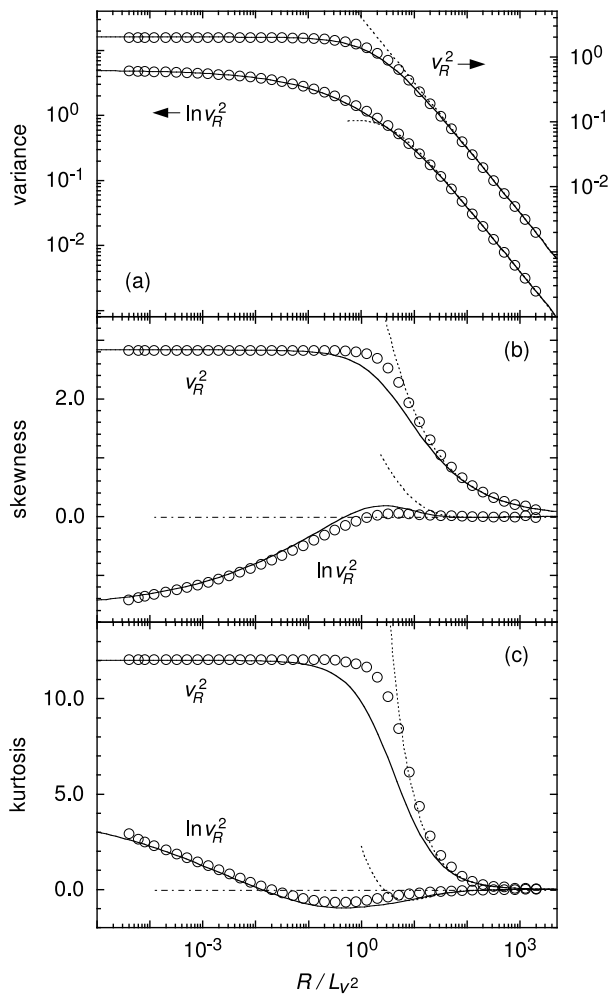}}
\caption{\label{f3} Variance (a), skewness (b), and kurtosis (c) of $\bar{v}^2_R$ and of $\ln \bar{v}^2_R$ as a function of $R/L_{v^2}$ for the random field of Eq.~(\ref{eq9}). The variance of $\bar{v}^2_R$ is normalized with the square of $\langle \bar{v}^2_R \rangle$. The solid and the dotted curves are model predictions  described in Sec.~\ref{S4} and in the Appendix. The dot-dashed lines denote the Gaussian value of $0$.}
\end{figure} 
%_________________________________________

Figure \ref{f3} compares the skewness and kurtosis of $\bar{v}_R^2$ with those of $\ln \bar{v}_R^2$ as a function of $R/L_{v^2}$ (open circles). With an increase in $R/L_{v^2}$, they decay from the values of $v^2$ or of $\ln v^2$ to the Gaussian value of $0$ (dot-dashed lines). The decay is much faster for $\ln \bar{v}_R^2$. Its skewness and kurtosis are already close to $0$ at $R/L_{v^2} \simeq 10^0$, where the skewness and kurtosis of $\bar{v}_R^2$ are yet close to the values of $v^2$. Thus, $\bar{v}_R^2$ tends to a log-normal distribution much faster than to any Gaussian distribution.

These features and others, i.e., features of the variances of $\bar{v}_R^2$ and of $\ln \bar{v}_R^2$, are reproduced by a model described in the Appendix (solid curves in Figs.~\ref{f2} and \ref{f3}). It is modelled that $\bar{v}_R^2$ at $R \gg L_{v^2}$ is determined by an additive process of an averaging over independent subsegments with length $4L_{v^2}$ \cite{m11}. Their number per segment with length $R$ is
%___________________________________________________________
\begin{subequations}
\label{eq11}
\begin{equation}
\label{eq11a}
N = \frac{R}{4L_{v^2}},
\end{equation}
%_________________________________________
which is the single parameter of the model. The gamma distribution of Eq.~(\ref{eq6}) is used for $z$ defined as
%___________________________________________________________
\begin{equation}
\label{eq11b}
z 
= 
\frac{N+1/\sqrt{2}}{\sqrt{2}}
\left( \frac{\bar{v}_R^2}{\langle \bar{v}_R^2 \rangle} - \frac{\sqrt{2}-1}{\sqrt{2}} \frac{N}{N+1/2} \right) > 0 ,
\end{equation}
%___________________________________________________________
with
%___________________________________________________________
\begin{equation}
\label{eq11c}
\gamma = \frac{(N+1/\sqrt{2})^2}{2N+1} .
\end{equation}
\end{subequations}
%___________________________________________________________
The distributions of $\bar{v}_R^2 / \langle \bar{v}_R^2 \rangle$ and of $z$ at each $R$ have the same shape. However, those of $\ln ( \bar{v}_R^2 / \langle \bar{v}_R^2 \rangle )$ and of $\ln z$ do not because a constant $(\sqrt{2}-1)/\sqrt{2} \times N/(N+1/2)$ has been added to $\bar{v}_R^2 / \langle \bar{v}_R^2 \rangle$. Actually among the values of $\bar{v}_R^2 / \langle \bar{v}_R^2 \rangle$ calculated for Fig.~\ref{f2}, only 0.0001\% of them are below the constant. This constant leads to a higher accuracy for $\bar{v}_R^2$ than for $z$ in an approximation as some multiplicative process (see Sec.~\ref{S2}). As a result, $\bar{v}_R^2$ in Fig. \ref{f3} is more log-normal than $z$ in Fig.~\ref{f1}.

%_________________________________________
\begin{figure}[bp]
\resizebox{8.8cm}{!}{\includegraphics*[4.9cm,11.5cm][17cm,26cm]{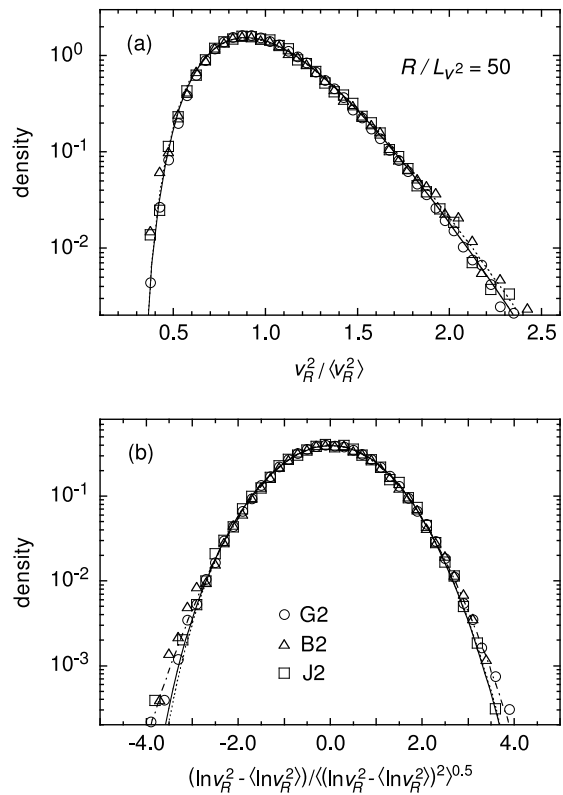}}
\caption{\label{f4} Same as in Fig.~\ref{f2} but for the data of a turbulent grid flow G2 (circles), a turbulent boundary layer B2 (triangles), and a turbulent jet J2 (squares) obtained in our past experiments using a wind tunnel \cite{m11}.}
\end{figure} 
%_________________________________________

We could observe a large-scale log-normal distribution for any homogeneous field of positive random variable, if Eq.~(\ref{eq5}) holds at each position, and also if any of the $n$-point correlations decays fast enough for the corresponding nonlinearity to become negligible at the large scales. The functional form of the correlation only affects details of the approach to the log-normal distribution. To normalize the scale $R$, we generalize the definition of the correlation length $L_{v^2}$ \cite{m11},
%_________________________________________
\begin{equation}
\label{eq12}
L_{v^2} 
= 
\frac{\int^{\infty}_0 \langle [v^2(x+r) - \langle v^2 \rangle][v^2(x) - \langle v^2 \rangle] \rangle \, dr}
     {2\langle v^2 \rangle^2} .
\end{equation}
%_________________________________________
Figure~\ref{f4} shows that $\bar{v}_R^2$ at $R \gg L_{v^2}$ in a variety of turbulent flows is log-normal \cite{m09} and is consistent with our model \cite{m11}, regardless of the correlation functions that depend on the configuration of the flow. Not so consistent with our model is $\bar{v}_R^2$ at $R \ll L_{v^2}$ (not shown here). This is because $v$ at each position $x$ is not exactly Gaussian. Also, no fluctuations exist below the scale of the Kolmogorov length \cite{k41b}, which is not $0$ so far as the Reynolds number is finite.

\section{Concluding Remarks} \label{S5}

The central limit theorem ensures that a sum of random variables is observed to become Gaussian with an increase in the total number of the variables \cite{ks77}, where the observation is just an approximation to ignore too large deviations of the sum from its average \cite{c38,t09}. However, if the distribution of each of the variables is positively skewed and satisfies the practical conditions of Eq.~(\ref{eq5}), the sum is observed to become log-normal before it is observed to become Gaussian. This is in contrast to the current consensus that any log-normal distribution is due to a multiplicative process, i.e., a product of random variables \cite{g30}, or equivalently to nonlinearity of the system. In fact, some are due to additive processes in linear systems that are divisible into independent subsystems. An analytical example is the gamma distribution of Eq.~(\ref{eq6}). Through Eq.~(\ref{eq11}), it reproduces the log-normal distribution observed for large-scale fluctuations of random scalar fields such as of turbulence \cite{m09}.

There have been studies similar to ours, i.e., studies for a sum of random variables that are exactly log-normal. No exact explicit formulae are known, but numerical calculations have shown that the sum is approximately log-normal \cite{f60,sy82}. This serves as a result for a special class of additive processes and is consistent with our result.

While we have discussed that a log-normal distribution is observable for a sum of positively skewed variables, the same discussion implies that a Gaussian distribution is observable for a product of negatively skewed variables. That is, Gaussianity is not necessarily due to an additive process in a linear system.

Thus far, we have focused on the cases where the central limit theorem holds because the variances of the summands are finite. If their distributions have, say, power-law tails of $z^{-1-\alpha}$ with $0 < \alpha < 2$ in the limit $z \rightarrow \infty$, the variances are infinite and the theorem does not hold. Then, the sum tends to a non-Gaussian stable distribution \cite{ks77,m82}. Its tail remains in the form of $z^{-1-\alpha}$. There is no notion of large deviations because they have to be much larger than the square root of the variance.

The log-normal distribution has been often related with such a power-law distribution \cite{m04}, by assuming that both of them arise from nonlinearity of the system. Indeed, the standard process for the log-normal distribution, i.e., multiplication of random variables, leads instead to a power-law distribution if a minimum value is set for the variables \cite{c53} or if random noise is added at each step of the multiplication \cite{k73}. These two distributions are nevertheless distinct. Although the log-normal distribution could have a long tail, all of its cumulants are finite. The power-law distribution has some infinite cumulants as shown above for the case of variance.

Having observed that the log-normal distribution also arises from additive processes in linear systems, we are rather interested in its relation with the Gaussian distribution. Of particular interest is to reconsider the existing observations of the log-normal distribution \cite{lwsa01,k41a,f57,t61,k62,ms79,h34,cp72,mhlm94,o62,bc03,m09}. The observed skewness is often small so that the distribution is confusingly similar to a Gaussian distribution \cite{lwsa01}. Such a case might turn out to be an example for a log-normal distribution from a sum, which tends to a Gaussian distribution with a further increase in the total number of the summands. The promising approach is to study the distribution beyond the range of the central limit theorem, i.e., large deviations from the average, or to study the process itself, e.g., whether it is for a linear or for a nonlinear system.

\begin{acknowledgments}
This work was supported in part by KAKENHI Grants No. 22540402 and No. 25340018. The author is grateful to T. Matsumoto for stimulating discussions.
\end{acknowledgments}

\appendix*
\section{Model of Equation (\ref{eq11})} \label{app}

The model of Eq.~(\ref{eq11}) is an extension of our past model \cite{m11}. We begin by obtaining the variance of $\bar{v}_R^2$ among the segments with length $R$ \cite{r54},
%_________________________________________
\begin{subequations}
\label{eqA1}
\begin{align}
\label{eqA1a}
&
\langle (\bar{v}_R^2 - \langle \bar{v}_R^2 \rangle )^2 \rangle \\
&
= \frac{2}{R^2} \int^R_0  
                          (R-r)
                          \langle [v^2(x+r) - \langle v^2 \rangle][v^2(x) - \langle v^2 \rangle] \rangle dr . \nonumber
\end{align}
%_________________________________________
If the two-point correlation of $v^2$ is negligible at large $r$, Eqs.~(\ref{eq12}) and (\ref{eqA1a}) yield
%_________________________________________
\begin{equation}
\label{eqA1b}
\langle (\bar{v}_R^2 - \langle \bar{v}_R^2 \rangle )^2 \rangle
\rightarrow
\frac{4L_{v^2}}{R} \langle v^2 \rangle^2 
\quad
\mbox{for}
\quad
R \rightarrow \infty.
\end{equation}
%_________________________________________
Thus, $L_{v^2}$ is naturally incorporated into our model. On the other hand, if the distribution of $v$ is Gaussian,
%_________________________________________
\begin{equation}
\label{eqA1c}
\langle (\bar{v}_R^2 - \langle \bar{v}_R^2 \rangle )^2 \rangle \rightarrow 2 \langle v^2 \rangle^2
\quad
\mbox{for}
\quad
R \rightarrow 0.
\end{equation}
%_________________________________________
These asymptotes are used to approximate Eq.~(\ref{eqA1a}) at any $R$ as
%_________________________________________
\begin{equation}
\label{eqA1d}
\langle (\bar{v}_R^2 - \langle \bar{v}_R^2 \rangle )^2 \rangle
\simeq
\frac{4 L_{v^2}}{R+ 2 L_{v^2}} \langle v^2 \rangle^2 .
\end{equation}
\end{subequations}
%_________________________________________
If the fluctuations of $\bar{v}_R^2$ could be regarded as in an equilibrium state, Eq.~(\ref{eqA1d}) has an analogue in the statistical mechanics, i.e., a formula for thermal fluctuations of energy $E$ in a canonical ensemble at temperature $T$ \cite{c85},
%_________________________________________
\begin{equation}
\label{eqA2}
\langle (E - \langle E \rangle )^2 \rangle = C T^2
\quad
\mbox{with}
\quad
C =  \frac{d \langle E \rangle}{d T} .
\end{equation}
%_________________________________________
This is the case even in turbulence \cite{m11}. We have defined the number of the subsegments with length $4L_{v^2}$ as $N = R/4L_{v^2}$ in Eq.~(\ref{eq11a}), which is used to relate Eq.~(\ref{eqA1d}) with Eq.~(\ref{eqA2}),
%_________________________________________
\begin{subequations}
\label{eqA3}
\begin{equation}
\label{eqA3a}
E = N \left[ \bar{v}_R^2 - ( 1-\sqrt{\zeta} ) \langle v^2 \rangle \right] + \frac{\bar{v}_R^2}{2},
\end{equation}
%_________________________________________
and
%_________________________________________
\begin{equation}
\label{eqA3b}
T = \frac{2N+1}{2\sqrt{\zeta}N+1} \langle v^2 \rangle .
\end{equation}
%_________________________________________
Here, $\zeta > 0$ is a constant. Then,
%_________________________________________
\begin{equation}
\label{eqA3c}
\langle E \rangle = CT 
\quad
\mbox{with}
\quad
C = \frac{(\sqrt{2\zeta} N+1/\sqrt{2})^2}{2N+1} .
\end{equation}
\end{subequations}
%_________________________________________
The first term in the right-hand side of Eq.~(\ref{eqA3a}) represents an additive process. This term is dominant over the second term in the thermodynamic limit of $N \gg 1$, where Eq.~(\ref{eqA3}) reproduces formulae of the thermodynamics \cite{m11}.

The energy distribution $f(E)$ in any canonical ensemble is determined by the heat capacity $C$ through a series of relations in the statistical mechanics \cite{c85}. From $C$ in Eq.~(\ref{eqA3c}), we obtain $f(E)$ in the form of the gamma distribution of Eq.~(\ref{eq6}),
%_________________________________________
\begin{subequations}
\label{eqA4}
\begin{equation}
\label{eqA4a}
f(E) = \frac{E^{C-1}\exp(-E/T)}{\Gamma(C) T^C} .
\end{equation}
%_________________________________________
Especially in the thermodynamic limit, the distribution is reduced to our past model \cite{m11},
%_________________________________________
\begin{equation}
\label{eqA4b}
f(E) = \frac{E^{\zeta N-1}\exp(-E/T)}{\Gamma(\zeta N) T^{\zeta N}}
\quad
\mbox{for}
\quad
N \gg 1 .
\end{equation}
\end{subequations}
%_________________________________________
This model corresponds to the exact Eq.~(\ref{eqA1b}) rather than to the approximate Eq.~(\ref{eqA1d}). By also using Eqs. (\ref{eq7a}) and (\ref{eqA3a}), we have $\langle (\bar{v}_R^2)^m \rangle_c \propto 1/N^{m-1}$ and hence $\propto 1/R^{m-1}$ in accordance with Eq.~(\ref{eq2b}).

To set the value of $\zeta$, we assume universality of $f(E)$ at $N \gg 1$. Since any interaction occurs only between adjacent members of the  subsegments, the energy distribution $f(E)$ at $N = R/4L_{v^2} \gg 1$ is determined by many random steps of the energy transfer from the subsegment length scale $4L_{v^2}$ to the segment length scale $R$. They should have randomized any effect of the individual subsegments \cite{k41b}, as implied by our results in Figs.~\ref{f2} and \ref{f4}. Consider at first a special case where the energies of the subsegments are independent of one another and are identically distributed with the distribution of the square of a zero-mean Gaussian random variable. The resultant $f(E)$ at $N \gg 1$ is the gamma distribution for $\gamma = N/2$ (see Sec.~\ref{S3}), which is described by Eq.~(\ref{eqA4b}) with
%_________________________________________
\begin{equation}
{\textstyle
\label{eqA5}
\zeta = \frac{1}{2}.
}
\end{equation}
%_________________________________________
Then, also in other general cases, the universality leads to $\zeta = 1/2$. This value reproduces the law of equipartition of energy $\langle E \rangle = NT/2$ \cite{c85} at $N \gg 1$ in Eq.~(\ref{eqA3c}). From Eqs.~(\ref{eqA3}), (\ref{eqA4a}), and (\ref{eqA5}), we obtain Eq.~(\ref{eq11}).

Lastly, we note that Eq.~(\ref{eqA4a}) does not serve as a very good model at the scales of $N = R/4L_{v^2} \simeq 1$ because we have not considered their details (see Fig.~\ref{f3}). The model could be refined by refining its approximation for $\langle (\bar{v}_R^2 - \langle \bar{v}_R^2 \rangle )^2 \rangle$ in Eq.~(\ref{eqA1d}). However, such refinement depends on the functional form of the two-point correlation, and hence it is not universal. The exact and universal model of Eq.~(\ref{eqA4b}) for $N \gg 1$ is rather reliable at least over a part of those scales (dotted curves in Figs.~\ref{f2}--\ref{f4}), where $\bar{v}_R^2$ has become almost log-normal.

\end{document}